\documentstyle[12pt,aasms4]{article}

\newbox\grsign \setbox\grsign=\hbox{$>$} 
\newdimen\grdimen \grdimen=\ht\grsign
\newbox\simlessbox \newbox\simgreatbox \newbox\simpropbox
\setbox\simgreatbox=\hbox{\raise.5ex\hbox{$>$}\llap
     {\lower.5ex\hbox{$\sim$}}}\ht1=\grdimen\dp1=0pt
\setbox\simlessbox=\hbox{\raise.5ex\hbox{$<$}\llap
     {\lower.5ex\hbox{$\sim$}}}\ht2=\grdimen\dp2=0pt
\setbox\simpropbox=\hbox{\raise.5ex\hbox{$\propto$}\llap
     {\lower.5ex\hbox{$\sim$}}}\ht2=\grdimen\dp2=0pt

\def\edcomment#1{\iffalse\marginpar{\raggedright\sl#1\/}\else\relax\fi}
\reversemarginpar

\begin{document}

\title{Chandra Grating Spectroscopy of the Seyfert Galaxy Ton~S180} 

\author {T.J.Turner \altaffilmark{1,2}, 
I.M. George \altaffilmark{1, 2}, T.Yaqoob \altaffilmark{1,3}, 
G. Kriss \altaffilmark{3, 4}, D.M. Crenshaw \altaffilmark{5}, 
S. Kraemer \altaffilmark{5}, 
  W. Zheng\altaffilmark{3}, 
J. Wang\altaffilmark{3}, 
K. Nandra\altaffilmark{1,6} 
}

\altaffiltext{1}{Laboratory for High Energy Astrophysics, Code 660,
        NASA/Goddard Space Flight Center,
        Greenbelt, MD 20771}
\altaffiltext{2}{Joint Center for Astrophysics, Physics Dept., University of Maryland
Baltimore County, 1000 Hilltop Circle, Baltimore, MD 21250} 
\altaffiltext{3}{Center for Astrophysical Sciences, Department of Physics and Astronomy, 
The Johns Hopkins University, Baltimore, MD 21218--2686}  
\altaffiltext{4}{Space Telescope Science Institute, 3700 San Martin Drive, Baltimore, MD 21218} 
\altaffiltext{5}{Catholic University of America; NASA/GSFC, Code 681, Greenbelt, 
MD 20771} 
\altaffiltext{6}{Universities Space Research Association }

\begin{abstract}

We present preliminary results from spectral 
observations of Ton~S180 using  {\it Chandra} and {\it ASCA}. The 
data confirm the presence of the soft excess but the {\it Chandra}  
LETG spectrum reveals it to be broad and smooth, rather than 
resolved into individual emission lines. This excess may represent 
either a primary or reprocessed continuum component or a blend of 
broad lines from an ionized accretion disk. 
The occurrence of a similar feature in five other NLSy1s leads us to 
conclude that 
this soft X-ray component may be a characteristic 
of sources accreting at a high rate. 

The X-ray spectrum shows no evidence
for absorption lines, indicating that if gas exists in the 
line-of-sight then it is in a very 
high ionization-state or has an extreme velocity distribution. 
The new {\it ASCA} data confirm that the narrow component of the Fe K$\alpha$ 
line peaks close to 
a rest-energy of 7 keV, indicating the presence of 
a significant amount of highly-ionized material in the nuclear environs.

\end{abstract}

\section{Introduction}

Examination of the Seyfert population shows that narrow-line Seyfert 1 
galaxies (NLSy1s) have relatively strong X-ray 
variability (Turner et al. 1999)   
and steeper X-ray spectra (Puchnarewicz et al. 1992; Laor et al 1994; 
Boller et al. 1996; Laor et al 1997; Brandt et al. 1997) 
than broad-line-Seyfert 1s (BLSy1s). 
A favored model explains NLSy1s as systems with relatively low mass 
black holes accreting at a high rate (e.g. Pounds, Done \& Osbourne 1995). 

Ton~S180  (PHL 912, z=0.06198; Wisotzki et al. 1995) is a bright NLSy1 with 
FWHM H$\alpha$ and FWHM H$\beta \sim 900 {\rm km\ s}^{-1}$. There is a  
low Galactic column density along the line-of-sight, 
$N_H=1.52 \times 10^{20} {\rm cm}^{-2}$ (Stark  et al. 1992). 
{\it BeppoSAX} (Comastri et al. 1998) and {\it ASCA} (Turner et al. 1998)
data indicated a photon index $\Gamma \sim 2.5$ in the 
2--10 keV band, and showed significant iron K$\alpha$ emission with a 
narrow peak at $\sim 7$ keV, suggesting the
circumnuclear material may be strongly ionized. 
Ton~S180 and another bright NLSy1, Ark~564,  show another noteworthy 
component,  unexplained  excess  emission close to 1 keV 
(e.g. Turner, George \& Netzer 1999, Turner, George \& Nandra 
1998; Vaughan et al. 1999, Comastri et al 2000).   

The first {\it Chandra} grating observations  
have resolved the complex soft X-ray spectra in some AGN, 
revealing absorption lines   
in two BLSy1s (Kaastra et al 2000; Kaspi et al. 2000a) with 
bulk outflow velocities several hundred km/s. 
The absorption lines have equivalent widths (EWs) of a few tens of 
m\AA\ and arise in gas with hydrogen column 
densities $\sim 10^{21} - 10^{22} {\rm cm^{-2}}$ 
consistent with those derived from absorption edges using {\it ASCA} data 
(Reynolds 1997, George et al 1998). 
Some X-ray line emission is also seen, consistent with an origin in 
the expanding shell of gas (Kaastra et al. 2000). 

A {\it Chandra} ACIS/LETG observation of Ton~S180 was performed as part of 
a multi-satellite campaign whose results 
will be presented in Turner et al. (2001; herafter T01)  and 
Edelson et al. (2001).  Here we concentrate 
on the {\it Chandra} and {\it ASCA} results, with reference to relevant 
results from contemporaneous  
{\it FUSE} and {\it HST} observations. 

\section{The X-Ray Data}
\subsection{The Chandra Data} 

There remain 
a number of calibration issues associated with LETG data. These 
make  
this presentation preliminary, however, the results and conclusions 
are thought to be 
robust to the expected refinements. Our simultaneous {\it ASCA}
data make it easy to determine where the problem areas are.  
The {\it Chandra} data were reprocessed using the latest calibration for 
the gain of the ACIS 
chips\footnote{\verb+ACISD1999-09-16GAINN0004.FITS+ 
supplied by the CXC}, 
and screened such that known bad pixels and columns were removed, as 
were events with detector `grades' {\it not} equal to 0,2,3,4 
or 6\footnote{See http://asc.harvard.edu/ciao/threads/}.
The light curve was examined and periods of high background excluded from 
the analysis.
Such screening resulted in an exposure of $\sim$75~ks.
The 1$^{st}$-order spectra were extracted from the screened event file
and ancillary response files constructed using 
{\tt CIAO} (v1.1.1). 

\begin{figure}[h]
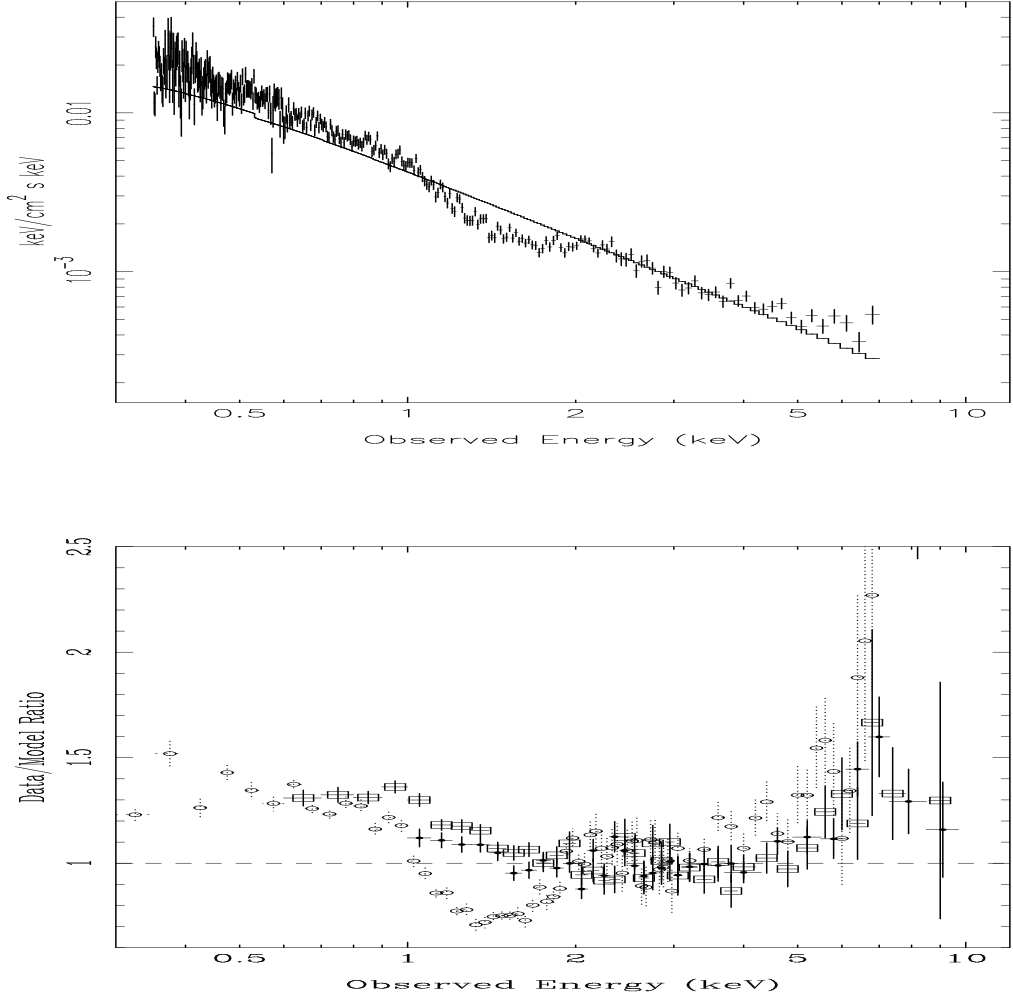

\plotfiddle{fig1a.vps}{6cm}{0}{75}{25}{-225}{+20}
\plotfiddle{fig1b.vps}{6cm}{0}{75}{25}{-225}{0}
\caption{
{\it Top Panel: 
The photon spectrum from LETG/ACIS data on Ton~S180. Positive and 
negative 1st-order data have been combined for the plot. 
The line represents a powerlaw, $\Gamma=2.44$ with 
Galactic absorption (based upon a fit to the 
1.5-5.0 keV plus 7.5-10.0 keV parts of the simultaneous {\it ASCA} data). 
The dip in the LETG data between 1 and 2 keV is 
due to inaccuracies in the current calibration.   
The soft excess is evident below 1 keV (12 \AA).
Bottom Panel. 
Data/model ratio for simultaneous {\it ASCA} 
and {\it Chandra} data, compared to 
the powerlaw continuum. Open ovals represent the combined first-order 
LETG data, the open squares are the combined SIS data and the crosses  
the combined GIS data. 
The spectra agree quite well except for the problematic 1-2 keV region where 
the maximum discrepancy is $\sim 40\%$. }
\label{fig:chandra_p1n1}  
}
 \end{figure}

Fig.~\ref{fig:chandra_p1n1}  shows the 1st-order LETG spectra of Ton~S180. 
A lack of strong spectral features is immediately obvious. 
The soft X-ray emission component discovered using 
{\it ASCA} (Turner et al. 1998) is confirmed, and the LETG data 
demonstrate that the spectrum is broad and smooth rather than 
resolved into discrete emission lines. 
This excess must be due to a
previously-unknown  primary or reprocessed continuum component or 
a blend of broadened  spectral features. 
In the absence of any detected features, it is difficult 
to place useful limits on the 
amount of absorbing gas. The resolution of the LETG corresponds to 
a FWHM $\simeq 1200 E_{keV} {\rm km/s}$. 
For a narrow line 
we find  90\% confidence limits on EW in the range 0.4--0.7 eV 
(7--22 m\AA) in the 0.5-1 keV 
regime, where 
absorption due to O{\sc vii} and O{\sc viii} might be expected. 
Assuming the velocity profile of the {\it FUSE} absorption lines, 
and using an appropriate curve of growth, we find an upper 
limit on $N$(O{\sc vii}) $\lesssim few \times 10^{16} {\rm cm^{-2}}$.   "

\subsection{ASCA}
{\it ASCA} observed Ton~S180 continuously for an 11-day period starting 
1999 December 3. For the full 
analysis of these data see T01. 
The subset of data which were simultaneous with {\it Chandra} 
show a powerlaw continuum 
of $\Gamma=2.44\pm0.04$, but  confirm the presence of   
the  soft excess in the 0.5--1 keV band 
\footnote{the {\it ASCA} analysis 
utilized an appropriate correction for the SIS detector 
degradation (Yaqoob, 2000)}.  
A steep powerlaw or bremmstrahlung component is an 
inadequate representation of this excess, as its form shows 
some curvature. 
However, the excess can be parameterized by a 
black body with rest-frame temperature 
$kT=158\pm4$ eV and absorption-corrected bolometric luminosity 
$L=1.2 \times 10^{44} {\rm erg\ s^{-1}}$. 
\begin{figure}[h]
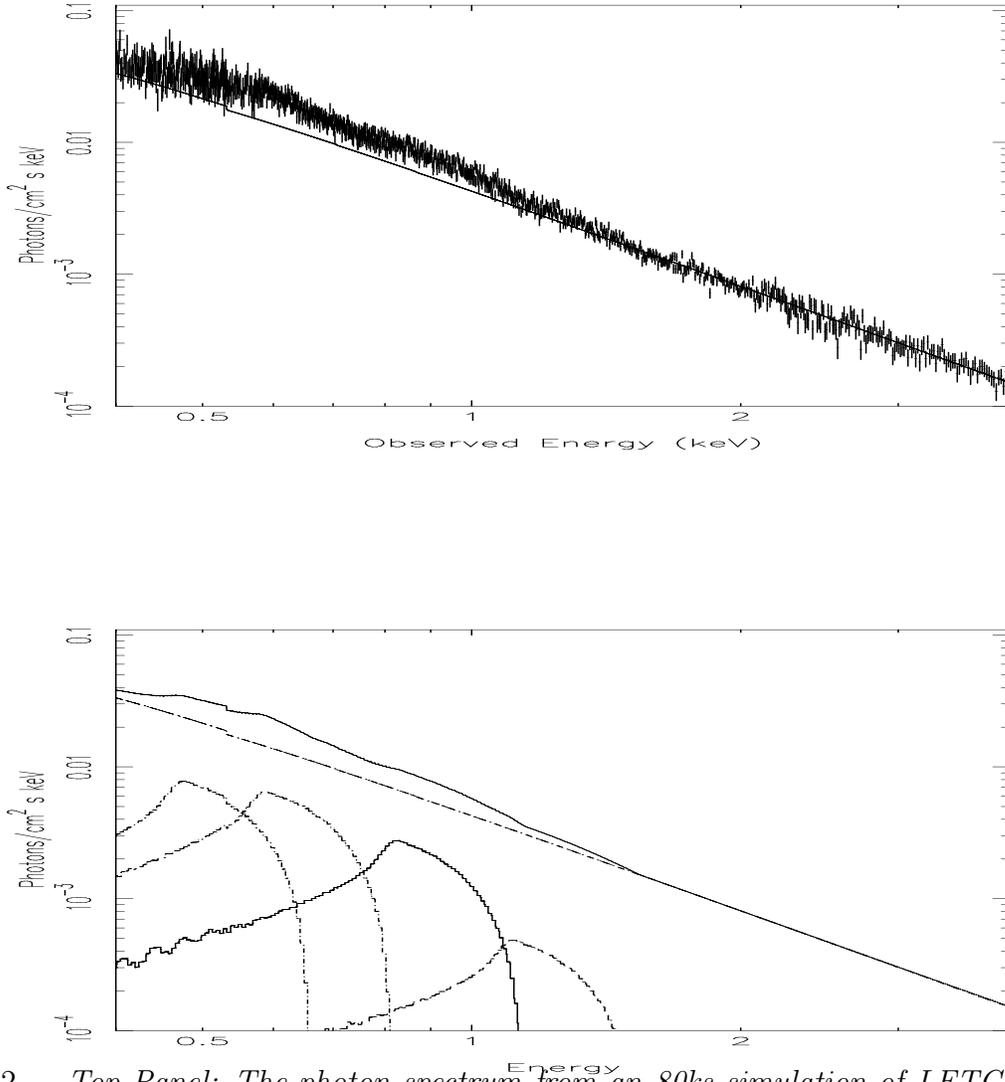

\plotfiddle{fig2a.vps}{6cm}{0}{75}{25}{-225}{+25}
\plotfiddle{fig2b.vps}{6cm}{0}{75}{25}{-225}{-25}
\caption{\it Top Panel: 
The photon spectrum from an 80ks simulation of LETG/ACIS data. The 
simulation shows the mean spectrum expected for the summed first-order data,
generated from a model which fits the {\it ASCA} data. 
The solid line shows only the powerlaw component of the model. 
It is clear that  
the sum of several soft X-ray lines from an accretion disk can produce a 
smooth observed excess of flux over the powerlaw. 
Bottom Panel: The model used for the simulation was a powerlaw with 
$\Gamma=2.44$ and  
Galactic absorption,  plus four broad emission components 
emitted from a Kerr accretion disk inclined at 
$45^{\rm o}$ to the line-of-sight. The inner and outer radii were  
2.4 $R_g$ and $ 400 R_g $, respectively. 
The disk emissivity was 
found to follow $ R^{-5.9} $ i.e. the total line profiles are strongly 
dominated by emission from the innermost regions of the disk. 
These parameters were determined using both the form of the 
soft excess and the broad Fe K$\alpha$ component observed 
in the {\it ASCA} spectra.
}
\label{fig:sim}  
\end{figure}
An alternative possibility is that the excess is  
the sum of a blend of broad lines from an ionized accretion disk. 
Simulations 
show that a wide range of parameter-space will result in lines 
sufficiently broad  that the horns of 
individual disk lines cannot be seen. 
For example, in Fig.2 we show how such a blend
can produce a broad feature inseparable from the continuum in LETG data.
Interestingly, Branduardi-Raymont et al (2000) find broad line emission 
from the innermost regions of a Kerr disk to describe the {\it XMM} 
RGS spectrum of two NLSy1s, Mrk~766 and MCG--$6-30-15$.  
Ballantyne, Iwasawa \& Fabian (2000) also find an ionized disk a good 
model for the X-ray spectra of five NLSy1s. In Ton~S180 
the soft component shows evidence for some variations in flux within the long
{\it ASCA} observation on timescales of $\gtrsim 200$ks 
suggesting a size $ < 6 \times 10^{15}$cm for the emitting region. There is
no obvious correlation between the flux of the excess and that of the 
hard X-ray continuum, and thus the EW of the soft component also shows
significant variability within the {\it ASCA} observation. 

The Fe K$\alpha$ line was parameterized using the full 11 days of data, as no
significant variability was observed in the line flux. 
The line profile was broad, 
asymmetric, and similar to that observed in a previous  {\it ASCA}
observation (Turner et al. 1998). The line could be  parameterized 
as the sum of a 
broad plus narrow Gaussian component. The rest-energy of the narrow line 
(fixed at 5 eV width) 
was $E_N=6.83^{+0.11}_{-0.16}$ keV, with EW
$88^{+31}_{-48}$ eV. The broad component gave $E_B=6.53^{+0.32}_{-0.28}$ keV, 
width FWHM $\sim 55,000$ km/s and EW 
$378^{+96}_{-136}$ eV, yielding an improvement to the fit 
$\Delta \chi^2=105$ for 1510 degrees of freedom ({\it dof}).   
The narrow line improved the fit at 
$ > 99$ \% confidence ($\Delta \chi^2=13$ for 1510 {\it dof}) when the broad 
line was modeled as a Gaussian, but no improvement 
when the broad line was modeled as a disk line. The energy of the narrow 
line is consistent with emission from Fe {\sc xxv-xxvi}, while the broad
line is consistent with ionization states Fe{\sc i} -Fe{\sc xxv}.  
Unfortunately it is not possible to simultaneously constrain the 
ionization-state and inclination of the disk. 

\section{Contemporaneous UV Observations}

Contemporaneous UV data 
taken with {\it FUSE} and {\it HST} will be presented in T01. 
Here we note some relevant results. 
{\it FUSE} data show absorption at three velocities near 
the redshift of Ton S180 
in the {\sc O~vi} $\lambda\lambda$1032,1038 resonance doublet.
 The {\sc O~vi} absorption line measurements 
reveal FWHM 75, 27 and 44 km/s for the three 
components, each with column density 
$N$(O{\sc vi})$\sim 10^{14}\ {\rm cm^{-2}}$. 
The UV absorbers are within 150 km/s of the systemic redshift and may 
arise in ionized interstellar material in the host galaxy.

An {\it HST} STIS spectrum 
with  a spectral resolving power of
$\lambda$/$\Delta\lambda$ $\approx$ 1000 showed no UV absorption lines.
Although we cannot rule out the possibility of
weak absorption at this resolution, we estimate an upper limit on the 
EW of any C{\sc iv}$\lambda$1550 absorption line to be 0.3 \AA\ 
(i.e. $N$(C{\sc iv})$\lesssim 2\times10^{13}\ {\rm cm^{-2}}$).

\section{Discussion}

\subsection{A New Continuum Component ?}

{\it ASCA} observations of Ton~S180 have shown a persistent
excess of  emission below $\sim 1.5$ keV relative to the hard X-ray 
continuum. It was previously suggested that 
this was due to unresolved line emission from
ionized species of Ne and from the Fe L-shell (Turner et al. 1998).
However, detailed analysis of a similar
X-ray feature in Akn~564 (Turner, George \& Netzer, 1999) 
showed that it was impossible to produce
enough line  emission from either thermal 
or photoionized gas, without producing other strong 
the features (e.g. Si, S etc) ruled out by the {\it ASCA} data. 
Another suggestion, based on PSPC data, was an origin in a 
curving continuum with  absorption features imprinted upon it 
(Brandt et al. 1994). Again, it was impossible to 
produce sufficiently strong {\it isolated} absorption features 
to explain the {\it ASCA} data. The new LETG spectrum now clarifies 
this issue, showing that the soft excess is not resolved into 
emission lines but is broad and smooth. Thus the soft excess 
emission must be primarily 
due to a continuum component (primary or comptonized) 
or a blend of very broad features, 
such as emission lines from the innermost regions of an ionized accretion disk.

The luminosity and temperature from the black body parameterization
 indicate emission from a region 
of remarkably small size, $\sim 10^{11}$cm in radius.
Using the preliminary spectral-energy-distribution 
from the multi-satellite data (T01) we estimate the bolometric luminosity 
of Ton~S180 to be $\sim 5 \times 10^{45} {\rm erg\ s^{-1}}$. 
Using this luminosity and the prescription of Laor (1998;  based upon 
luminosity and FWHM H$\beta$) we estimated the black hole mass in Ton~S180 
to be $\sim 10^7 {\rm M}_{\odot}$.  
However, the Eddington luminosity  is 
$L_{Edd}= 1.3 \times 10^{45} M/10^7M_{\odot} {\rm erg\ s^{-1}}$, so 
to avoid Ton~S180 exceeding the Eddington limit, we require 
the black hole mass to be $\gtrsim 4 \times 10^7{\rm M}_{\odot}$.
Two further independent estimates of the black hole mass can be made based 
upon the variability characteristic of the source (Laor 2000), and 
the luminosity at 5100 \AA\ (Kaspi et al. 2000b). Both of these methods yield 
 $M_{BH} \sim 8 \times 10^7 M_{\odot}$. The innermost stable orbit 
in this case is $\sim 7 \times 10^{13} {\rm cm}$, or 
$\sim 3 \times 10^{13} {\rm cm}$ for a maximally spinning hole.  
Thus our estimate 
of the radius of the emitting region appears inconsistent with the innermost 
edge of the accretion disk. This indicates a simple black body 
parameterization of the soft excess is probably inadequate. The possibility 
of a parameterization as a Comptonized black body will be addressed 
in T01. 
For black holes operating near $L_{Edd}$, the accretion disk surface is 
predicted to be highly ionized 
and the disk spectrum can produce a strong soft excess. Matt et al. (1993) 
show that for high accretion rates and  black hole masses typical of AGN,  
strong soft emission  
will be evident below 1 keV and Fiore et al. (1998) suggested 
an ionized disk as the origin of a soft X-ray component in 
PG $1244+026$ which is characteristically similar 
to that observed in Ton~S180. Simulations show that the broad 
lines expected from an ionized accretion disk 
can produce a smooth and broad excess of emission 
 which is consistent with the soft excess, Fe K$\alpha$ line 
and the {\it Chandra} grating data. 

Interestingly, the five AGN known to possess a soft spectral component 
similar to Ton~S180 are all NLSy1s 
 (Fiore et al. 1998, George et al. 2000, Turner et al. 1999). 
Thus we conclude that this particular type of soft X-ray component 
appears to be a characteristic  
of sources accreting at a high rate. Matt et al. (1993) demonstrate that the
spectrum of an ionized accretion disk is a strong function of the accretion
rate (and black hole mass), so there may be a critical
accretion rate at which the observed characteristics of the source change
significantly.

\subsection{The State of the Circumnuclear Gas}

A lack of absorption lines in Ton~S180 
means line-of-sight material is absent, highly ionized, 
or has turbulent velocities 
which are either extremely high (many thousand km/s) 
or significantly lower (few hundred km/s) than the resolution of the LETG.  
In the latter case, we would still be able to detect absorption edges, 
but none are evident. The 90\% confidence limit on 
O{\sc vii} (0.7393 keV) is $\tau < 0.16$ which places a  
limit of $N$(O{\sc vii}) $\lesssim 7\times10^{17} {\rm cm^{-2}}$. 
The upper limit from the absorption line, $N$(O{\sc vii}) 
 $\lesssim few \times 10^{16} {\rm cm^{-2}}$ 
clearly provides 
more than an order of magnitude tighter constraint on the column.  
Combining this limit with the measurement from {\it FUSE} 
$N$(O{\sc vi})$\sim 10^{14}\ {\rm cm^{-2}}$ yields a ratio 
O{\sc vii}/O{\sc vi} $\sim 250$, within the expected range 
for gas in photoionization equilibrium. 
Indeed, the gas is very likely
to be in a high state of ionization 
since (\S3) $A$(O/C)$N$(O{\sc vi})/$N$(C{\sc iv})$\gtrsim3$
(where $A$(O/C) is the relative abundance of O and C).
Some other NLSy1s show significant UV absorption (Crenshaw et al. 1999),  
thus it is unclear whether or not conditions in the circumnuclear gas 
in Ton~S180 are linked to the presence of the soft spectral component. 

The peak of the Fe K$\alpha$ line 
is at $\sim 7$ keV. If this represents a separable
narrow line then the EW implies 
emission from a full shell of material with 
$N_H \sim 6 \times 10^{22}{\rm cm^{-2}}$, 
using Leahy \& Creighton (1993) estimates for neutral material 
adjusted to a mix of Fe {\sc xxv-xxvi}. The 
ionization state is large enough to allow production of the 
narrow line in the line-of-sight without 
a significant soft X-ray opacity. However, the presence of the broad 
component indicates a significant amount of reprocessing 
material out of the line-of-sight.

\section{Summary}

The {\it Chandra} ACIS/LETG spectrum shows that the soft excess in Ton~S180
is smooth and broad and therefore is not due to a blend of 
individually narrow emission lines from
photoionized or collisionally-ionized gas. The soft excess 
could be a primary or reprocessed continuum component. 
A simple black body can be ruled out on the basis of the inferred 
size of the emission region. 
The occurrence of a similar 
feature in several NLSy1s indicates that 
this type of soft X-ray component 
may be a characteristic 
of sources accreting at a very high rate.
A blend of very broad emission 
lines from the inner regions of an ionized accretion disk is 
consistent with the 
{\it ASCA} and {\it Chandra} data. 

The {\it Chandra} LETG spectrum shows no evidence for line-of-sight 
absorption. High-resolution  UV spectra show only a low column of 
highly-ionized gas, consistent with the upper limit 
on the strength of absorption lines in the soft X-ray band. 
The narrow component of the 
Fe K$\alpha$ emission line could be consistent with  
highly-ionized gas in the line-of-sight, however, a strong broad component 
shows that 
there must also be a significant amount of  circumnuclear material out 
of the line-of-sight.

\section{Acknowledgements}

We are grateful to {\it Chandra} team for their operation of the satellite 
and especially to Herman Marshall and Dave Huenemoerder for discussions 
regarding calibration and data analysis. This research has 
made use of data obtained through the
HEASARC on-line service, provided by NASA/GSFC.
This work is based in part on data obtained for the
Guaranteed Time Team by the
NASA-CNES-CSA {\it FUSE} mission operated by the Johns Hopkins University.
Financial support to U. S. participants has been provided by
NASA contract NAS5-32985.
G. Kriss acknowledges additional support from NASA 
LTSA grant NAGW-4443. 
T.J.Turner acknowledges support from  LTSA grant NAG5-7538. D.M.Crenshaw 
and S.Kraemer acknowledge support from NAG5-4103.

\end{document}